\documentclass[%
 reprint,
superscriptaddress,
aps,
prb,
]{revtex4-2}

\usepackage{graphicx}
\usepackage{dcolumn}
\usepackage{bm}
\usepackage{siunitx}
\usepackage{amsmath}
\usepackage{amssymb}
\usepackage{gensymb}
\usepackage{multirow}
\usepackage{verbatim}
\usepackage{ulem}
\usepackage{hyperref}
\hypersetup{colorlinks=true, citecolor=blue, urlcolor=black, linkcolor=blue}
\usepackage{subfiles}
\usepackage{ccaption}
\usepackage{caption}
\setcitestyle{super}
\usepackage{titlesec}
\titleformat{\section}
  {\normalfont\normalsize\bfseries\centering}  
  {\thesection}                                
  {1em}                                        
  {}                                          

\begin{document}
\title{Chip-scale ultrafast soliton laser}
\author{Qili Hu}
\affiliation{Institute of Optics, University of Rochester, Rochester, NY, USA}
\author{Raymond Lopez-Rios}
\affiliation{Institute of Optics, University of Rochester, Rochester, NY, USA}
\author{Zhengdong Gao}
\affiliation{Department of Electrical and Computer Engineering, University of Rochester, Rochester, NY, USA}
\author{Jingwei Ling}
\affiliation{Department of Electrical and Computer Engineering, University of Rochester, Rochester, NY, USA}
\author{Shixin Xue}
\affiliation{Department of Electrical and Computer Engineering, University of Rochester, Rochester, NY, USA}
\author{Jeremy Staffa}
\affiliation{Institute of Optics, University of Rochester, Rochester, NY, USA}
\author{Yang He}
\affiliation{Department of Electrical and Computer Engineering, University of Rochester, Rochester, NY, USA}
\author{Qiang Lin}
\thanks{Correspondence to qiang.lin@rochester.edu}
\affiliation{Institute of Optics, University of Rochester, Rochester, NY, USA}
\affiliation{Department of Electrical and Computer Engineering, University of Rochester, Rochester, NY, USA}

\begin{abstract}
Femtosecond laser, owing to their ultrafast time scales and broad frequency bandwidths, have substantially changed fundamental science over the past decades, from chemistry and bio-imaging to quantum physics. Critically, many emerging industrial-scale photonic technologies—such as optical interconnects, AI accelerators, quantum computing, and LiDAR—also stand to benefit from their massive frequency parallelism. However, achieving a femtosecond-scale laser on-chip, constrained by size and system power input, has remained a long-standing challenge.
Here, we demonstrate the first on-chip femtosecond laser, enabled by a new mechanism---photorefraction-assisted soliton (PAS) mode-locking. Operating from a simple, low-voltage electrical supply, the laser provides deterministic, turn-key generation of sub-90-fs solitons. Furthermore, it provides electronic reconfigurability of its pulse properties and features an exceptional optical coherence with a 53\,Hz intrinsic comb linewidth. This demonstration removes a key barrier to the full integration of chip-scale photonic systems for next-generation sensing, communication, metrology, and computing.

\end{abstract}

\maketitle

\section*{Introduction}

Femtosecond laser, which produce femtosecond pulses in the time domain while simultaneously forming broadband coherent optical frequency combs in the spectral domain, have become a foundational technology in modern science\cite{keller2003recent,cundiff2003colloquium,diddams2020opticala}. Their unique capabilities, including ultrafast time scales, massive frequency parallelism, and unparalleled frequency precision, underpin broad cross-discipline applications developed, from molecular chemistry\cite{zewail2000femtochemistry} and bio-imaging\cite{xu2013recent} to communications\cite{torres-company2019laser}, precision sensing\cite{coddington2016dualcomb,picque2019frequencya}, and optical clocks\cite{ludlow2015optical}.
The impact of these ultrafast broadband lasers extends beyond fundamental science in the laboratory, promising to transform emerging industry-scale photonic technologies that rely on multi-wavelength sources, such as optical interconnects\cite{hsia2023integrated,rizzo2023massivelya}, autonomous driving\cite{riemensberger2020massively}, 5G/6G communication\cite{kudelin2024photonicc,sun2024integrateda}, and AI accelerators\cite{ahmed2025universal,hua2025integrated}. Realizing femtosecond-scale broadband combs on-chip, within constrained space and power budgets, is critical for enabling the system parallelism, energy efficiency, and integration required to make these applications practical and mass-producible.

\begin{figure*}[htbp]
\centering
\includegraphics[width=\textwidth]{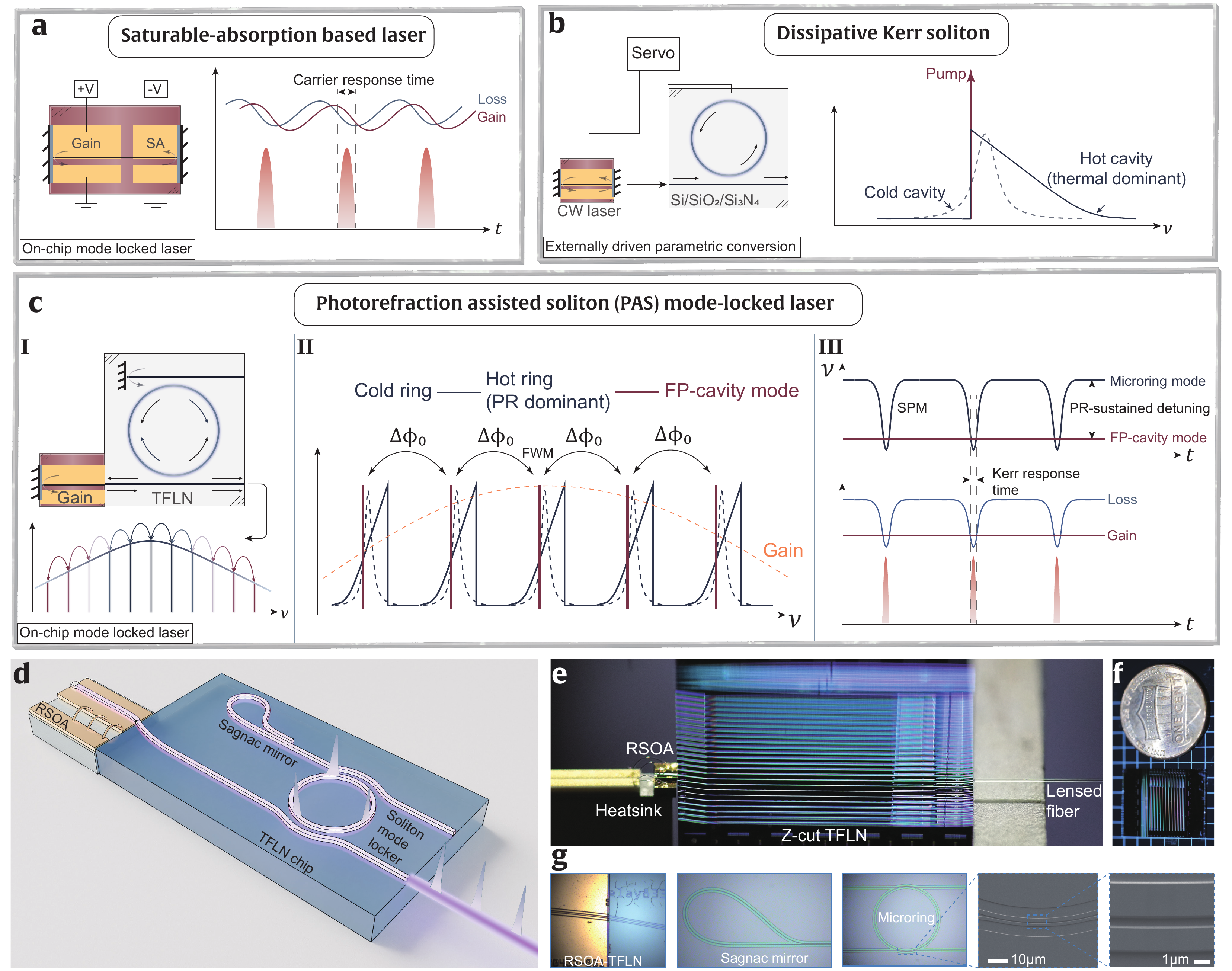}
\caption{\textbf{Concept and implementation of on-chip soliton laser.} 
\textbf{(a, b)}~state-of-the-art approaches to chip-scale ultrashort pulse generation.
\textbf{(a)}~On-chip MLLs. Typically based on III-V platforms, these devices use a reverse-biased semiconductor saturable absorber to achieve self-starting pulse generation directly from an electrical current. However, their pulse duration is fundamentally limited to the sub-picosecond domain by the finite carrier response time of the semiconductor SA. 
\textbf{(b)}~Externally-driven DKS. A high-Q microresonator is pumped by an external continuous-wave (CW) laser to have solitons through four-wave mixing (FWM). While this approach generates femtosecond pulses, the externally-driven parametric process is critically reliant on the external pump laser, introducing challenges such as difficult soliton initiation, low conversion efficiency, pedestal-present soliton, and pump-dependent comb linewidth.
\textbf{(c)}~Chip-scale femtosecond soliton laser enabled by PAS mode-locking. \textbf{(c.I)}~Device architecture. A microring exhibiting a PR effect is embedded within a semiconductor-gain-based FP external cavity. \textbf{(c.II)}~Frequency-domain picture of the soliton mode-locking mechanism. \textbf{(c.III)}~Time-domain behavior of the mode-locking.  
\textbf{(d)}~Schematic of the hybrid integrated laser architecture.
\textbf{(e,f)}~Photograph showing the hybrid integration and the device footprint in comparison with a one-cent coin.
\textbf{(g)}~Micrographs detailing the RSOA-TFLN coupling interface, embedded high-Q microresonator, and broadband Sagnac mirror.}
\label{Fig1}
\end{figure*}

However, realizing femtosecond-pulse source on photonic chips remains fundamentally challenging. Current on-chip mode-locked lasers (MLLs) based on semiconductor platforms (Fig.\,\ref{Fig1}a) offer excellent scalability and, crucially, spontaneous soliton generation from a simple direct-current bias\cite{davenport2018integratedb,wei2022advancesa,rafailov2007modelocked}. However, their pulse performance is fundamentally constrained by the finite carrier relaxation times inherent to saturable absorption (SA) mode-locking, leading to sub-ps scale at best. Alternatively, externally-driven dissipative Kerr solitons (DKS) in high-Q microresonators (Fig.\,\ref{Fig1}b) can achieve femtosecond-scale pulses and high comb line coherence\cite{kippenberg2018dissipative,pasquazi2018microcombs}, and have been applied to many areas\cite{sun2023applications}. However, DKS is essentially not a self-contained laser and relies critically on external laser pumping for proper operation. This external-driving scheme introduces fundamental challenges including operational complexity, pump residue, and power inefficiency. Current solutions to these issues generally sacrifice the core benefits of integration by requiring complex and bulky external control elements or sophisticated fiber components for the tuning, stabilization, and synchronization of frequency and time
\cite{shen2020integrateda,wang2020advancesa,xue2019superefficient, helgason2023surpassinga,bruch2021pockels,bao2019laserb,nie2022dissipative}, leading to significant system complexity and substantial additional power budget. To date, a fully integrated, power-efficient on-chip light source capable of generating femtosecond-scale pulses has remained elusive.

In this work, we demonstrate the first chip-scale femtosecond laser. This on-chip soliton laser is based on a new mechanism: photorefraction-assisted soliton (PAS) mode-locking. As a result, the laser achieves pulse durations shorter than 90\,fs with a repetition rate up to 3\,THz. Operating from a simple low-voltage electrical supply, this on-chip laser exhibits deterministic, turn-key soliton generation and outstanding long-term stability. Moreover, the device is electronically reconfigurable, allowing its repetition rate and spectral bandwidth to be tuned via the pump current. The laser also exhibits outstanding long-term stability and a 53-Hz intrinsic comb linewidth, a coherence three orders of magnitude better than the current state-of-the-art for on-chip MLLs.

\begin{figure*}[htbp]
\centering
\includegraphics[width=\textwidth]{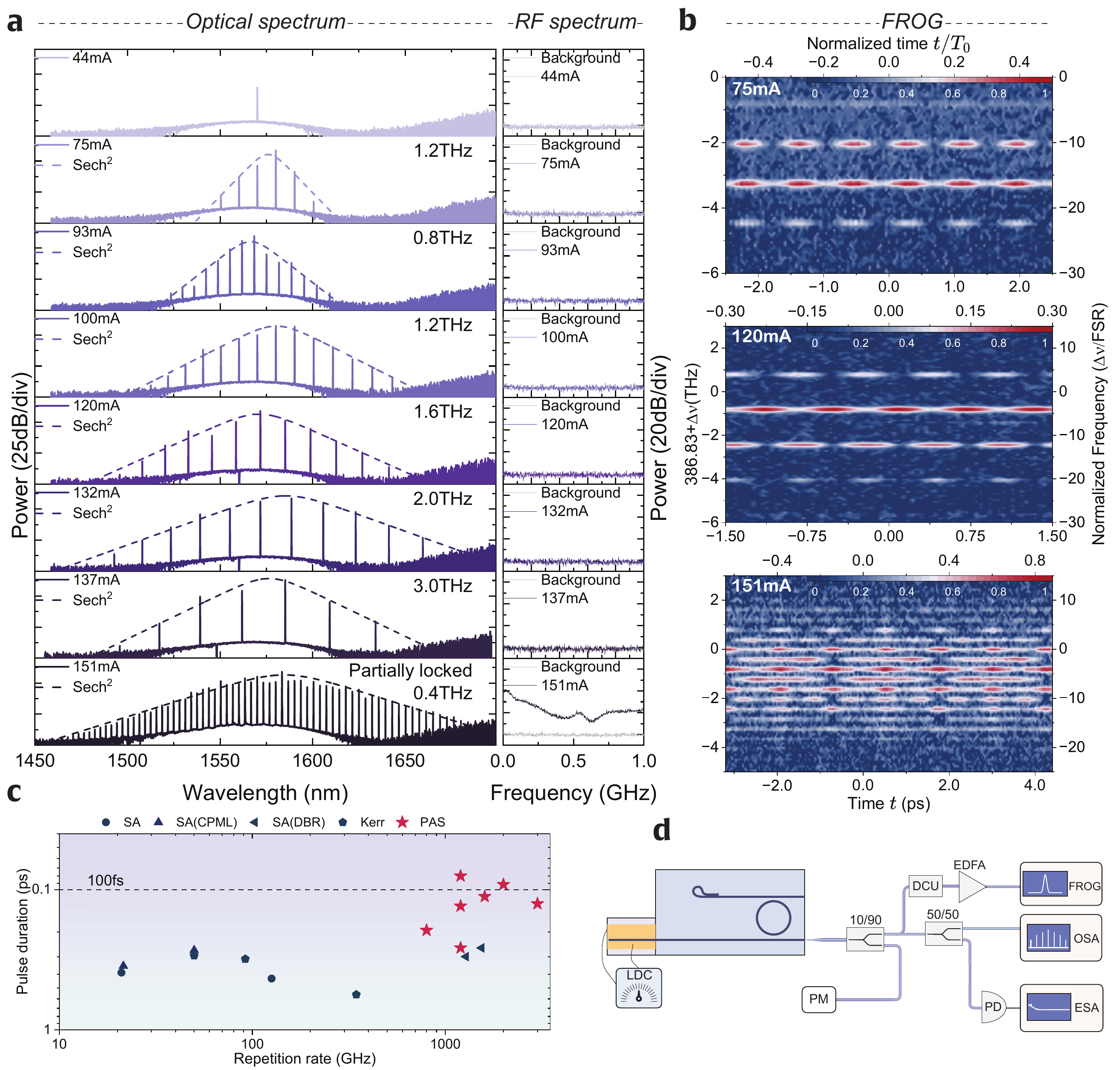}
\caption{\textbf{Performance of the soliton microcomb laser.} 
\textbf{(a)}~Optical (left) and RF (right) spectra of laser output at different RSOA driving currents. Dashed lines represent $sech^2$ fits. Each state was obtained by optimizing the relative position between the RSOA and TFLN chips to compensate for thermal-induced mechanical drift. 
\textbf{(b)}~FROG traces for the 1.2-THz (top), 1.6-THz (middle), and 0.4-THz (bottom) soliton states at RSOA driving currents of 75, 120, and 151~mA, respectively. The center frequency ($\Delta \nu=0$) of the FROG traces is 386.83THz (corresponding wavelength of 775~nm). For a better comparison, the figures also show frequency and time scales normalized, respectively, by the 200-GHz FSR and 5-ps roundtrip time ($T_0$) of the mode-locking resonator. The relatively indistinct pulse temporal boundaries in the FROG traces result simply from limited gain bandwidth of the EDFA used to boost the soliton pulses before they are recorded by FROG (Fig.\,\ref{Fig2}d). This artifact is confirmed by performing the same FROG on similar comb states produced via conventional DKS excitation in a passive microresonator. Details are provided in SI. 
\textbf{(c)}~Comparison with advanced on-chip MLLs on pulse duration and repetition rate. All data are listed in Table\,S1 of SI. The 132\,mA state is 90\,fs and 2\,THz. The 80\,fs state is given later in Fig.\,\ref{Fig3}e.
\textbf{(d)}~Schematic of experimental setup. LDC: laser diode controller; DCU: dispersion compensation unit; PM: power meter; FROG: frequency-resolved optical gating; OSA and ESA: optical and electrical spectrum analyzer. }. 
\label{Fig2}
\end{figure*}

\section*{Laser structure and operation principle}

Our laser integrates an InP‑based reflective semiconductor optical amplifier (RSOA) as the gain medium with a thin-film lithium-niobate (TFLN) photonic integrated circuit (PIC) as the external soliton laser cavity (Fig.\,\ref{Fig1}c.I). TFLN PICs have recently been shown as a promising platform for a wide range of electro-optic, nonlinear photonic, quantum photonic, and laser applications \cite{zhu2021integrateda, lin2020advances, boes2023lithium}. It exhibits a unique photorefractive (PR) effect that can self-stabilize the optical resonances of a high-Q TFLN microresonator in the red-detuned regime \cite{he2019selfstarting}. We thus utilize this intriguing feature inside a laser cavity for turn-key mode locking operation as illustrated in Fig.\,\ref{Fig1}c. 

The laser comprises three key elements (Fig.\,\ref{Fig1}d): InP-based RSOA as the optical gain, a TFLN microring resonator as the Kerr mode-locking element, and a Fabry–Pérot (FP) cavity as the principal laser cavity formed by a broadband Sagnac end mirror on the TFLN PIC side and a reflective end mirror at the facet of the InP RSOA. In this architecture, the FP cavity primarily determines the lasing mode frequencies, while the embedded high-Q microresonator governs the nonlinear optical dynamics.

Upon applying an injection current to the RSOA, multimode lasing waves initiated from the III-V optical gain will pass through the embedded high-Q microresonator, which produces a strong PR effect to blue-shift the optical resonances of the microresonator. This process intrinsically stabilizes the system by forcing the FP cavity modes to operate on the red-detuned side of the microresonator's resonances (Fig.\,\ref{Fig1}c.II). Simultaneously, the strong optical Kerr effect inside the microresonator converts these lasing waves into a broadband frequency comb via the resulting four-wave mixing process. As the comb is produced in the stable red-detuned regime ideal for soliton formation, it would naturally evolve into a mode-locked soliton pulse train.

The PAS mode-locking can also be seen clearly in the time domain, as illustrated in Fig.~\ref{Fig1}c.III. Self-phase modulation (SPM) induced by optical pulses dynamically shifts the instantaneous microring resonance (blue line) toward the FP cavity mode (red line). This transient, self-regulating mechanism creates a net-gain window on an ultrafast timescale, favoring femtosecond pulse formation. The PR effect counteracts thermal nonlinearities within the microring mode-locker, which is essential for establishing the stable red-detuned condition required for this ultrafast Kerr modulation. PAS mode-locking can thus be understood as a form of Kerr mode-locking, a mechanism previously proven to be highly effective for realizing femtosecond lasers in free-space and fiber systems\cite{keller2003recent,lau2025sub100fs}. This work represents the first demonstration of Kerr mode-locking on-chip.

\section*{Laser characteristics and soliton comb dynamics}

Fig.~\ref{Fig1}d-f show an example of the fabricated laser device. The high‑Q TFLN microring features an instrinsic optica Q of 2.7 million (See Supplementary Information (SI), Fig.\,S2a) and a free spectral range (FSR) of 200\,GHz, about 77 times of that of the principal FP cavity (FSR of $\sim$2.6~GHz). To characterize the laser performance, the laser output is monitored through the bus waveguides coupled to the microring (designed to be over coupled, with a loaded optical Q of $\sim$1~million). We examined the optical and RF spectra across various operating states by systematically varying the RSOA current while manually optimizing the coupling position to compensate for thermal-induced misalignment between the gain chip and TFLN chip. 

The laser begins CW lasing at a relatively low threshold of 44\,mA, indicating a small cavity loss. Beyond 75\,mA, the laser transitions into various comb states, with comb mode spacings of (0.8$-$3)~THz when the RSOA current increases from 75 to 137~mA. As shown in Fig.\,\ref{Fig2}a, all these comb states exhibit characteristic $sech^2$ spectral profiles, indicating mode-locked soliton states. They correspond to the $N\times {\rm FSR}$ perfect soliton crystal states of the mode-locking microresonator in which the $N$ soliton pulses are spaced equally in time within one round trip \cite{cole2017soliton, karpov2019dynamics}. The RF signals detected from the soliton pulse train exhibit exceptionally clean spectrum approaching the background level, clear evidence of stable soliton mode locking. As the repetition frequencies of (0.8$-$3)~THz correspond to (4$-$15)$\times$FSR of the mode-locking microresonator (and $\sim$(308$-$1155)$\times$FSR of the principal FP laser cavity), the laser operates in the intriguing regime of high-order harmonic mode-locking. 

As shown in Fig.\,\ref{Fig2}a, the spectrum of the soliton combs broadens considerably with increased RSOA current, reaching a 3-dB optical bandwidth of $\sim$3.4\,THz when the current increases to 132\,mA. This spectral bandwidth corresponds to a pulse width as small as $\sim$90\,fs, significantly shorter than that can be achieved in a conventional on-chip MLL~\cite{hermans2022onchip, chang2022integrated}. Moreover, the recorded soliton comb spectrum closely matches the spectral characteristics of DKS states produced in a separate passive microring with identical dispersion design and similar coupling conditions, which further validates the soliton nature of these mode-locked comb states. The details are provided in the SI. 

To further verify the mode-locked nature of the soliton emitted from the laser, we examined their temporal characteristics using frequency-resolved optical gating (FROG). As shown in Fig.\,\ref{Fig2}b, FROG traces at 75\,mA and 120\,mA reveal highly periodic structures in both temporal and frequency domains (see SI for detailed discussion of FROG traces). For example, the FROG trace at 75\,mA shows an interference pulse train with a temporal period of $\sim$0.83~ps that corresponds to one-sixth of the fundamental 5-ps round-trip time of the mode-locking microresonator and a frequency spacing of 1.2~THz that corresponds to six times the fundamental 0.2-THz FSR. The FROG trace at 120\,mA shows similar behavior while with different temporal period of 0.625\,ps and frequency spacing of 1.6~THz. These observations clearly verify the coherent nature of the soliton comb in both time and spectral domains. Moreover, the clean FROG traces indicate that the soliton pulse trains are fully background-free, in stark contrast to DKS that is inevitably accompanied with a strong pump wave background. This can be seen more clearly by comparison with the FROG traces of DKS produced in a passive microresonator, whose details are provided in SI.

When the RSOA current increases to 151\,mA, the microcomb spectrum is further broadened with a comb mode spacing of 0.4\,THz. FROG characterization reveals a periodic but complex temporal and frequency structure, implying a partially mode-locked soliton state. The complex FROG trace infers a dual-pulse pattern with a period of 2.5\,ps that consists of a stable primary pulse and an unstable satellite pulse. The satellite pulse is likely unstable in phase and relative temporal position, manifesting in the FROG trace as pulse peaks residing on a temporally broadened pedestal. The instability of the satellite pulse leads to an elevated level in the detected RF spectrum. The intriguing unstable double-pulsing behavior is reminiscent of the over-pumping conditions or insufficient mode locking (e.g., small reverse bias in SA in conventional on-chip MLLs \cite{dong2023broadbanda,huang2024quantuma} as well as soliton rain in fiber soliton lasers \cite{chouli2010solitona}). A detailed investigation of this phenomenon lies beyond the scope of this work and will be explored in the future.

The comb power is measured to be about 1.4\,mW on chip with a RSOA driving current of 132\,mA. Given that the laser has two output ports (related to two bus waveguides coupled to the microring shown in Fig.\,\ref{Fig1}), the total output comb power is about 1.5\,mW. It corresponds to a wall-plug efficiency of $\sim$1.1\% that is on par with conventional on-chip MLLs~\cite{hermans2022onchip, chang2022integrated}, while here the soliton laser exhibits markedly broader 3-dB optical bandwidth (shorter pulse width) and significantly higher repetition rate. Fig.~\ref{Fig2}c compares this laser with state-of-the-art on-chip MLLs, which clearly shows the superior pulse performance beyond the reach of conventional on-chip MLLs. The enhanced performance directly arises from the ultrafast soliton mode‑locking mechanism and the effectively unlimited parametric bandwidth supported by design. The comb power is also on par with those produced with DKS approaches~\cite{hermans2022onchip, chang2022integrated}, while here the soliton laser requires an overall electric power orders of magnitude lower. Interestingly, we estimate the circulating optical power inside the principal FP cavity to be only around 3\,mW, about an order of magnitude lower than the optical power required to excite similar soliton combs under conventional DKS schemes (see SI, Fig.\,S2). This enhanced efficiency stems from the multi-mode pumping mechanism, where the soliton threshold is inversely proportional to the number of coherent in-phase excitation modes~\cite{obrzud2017temporala}. In the device, the broadband laser gain simultaneously supports multiple lasing modes, effectively creating a natural multi-mode self-pumping for the embedded mode-locking microring resonator (Fig.\,\ref{Fig1}).

\begin{figure*}[htbp]
\centering
\includegraphics[width=1\textwidth]{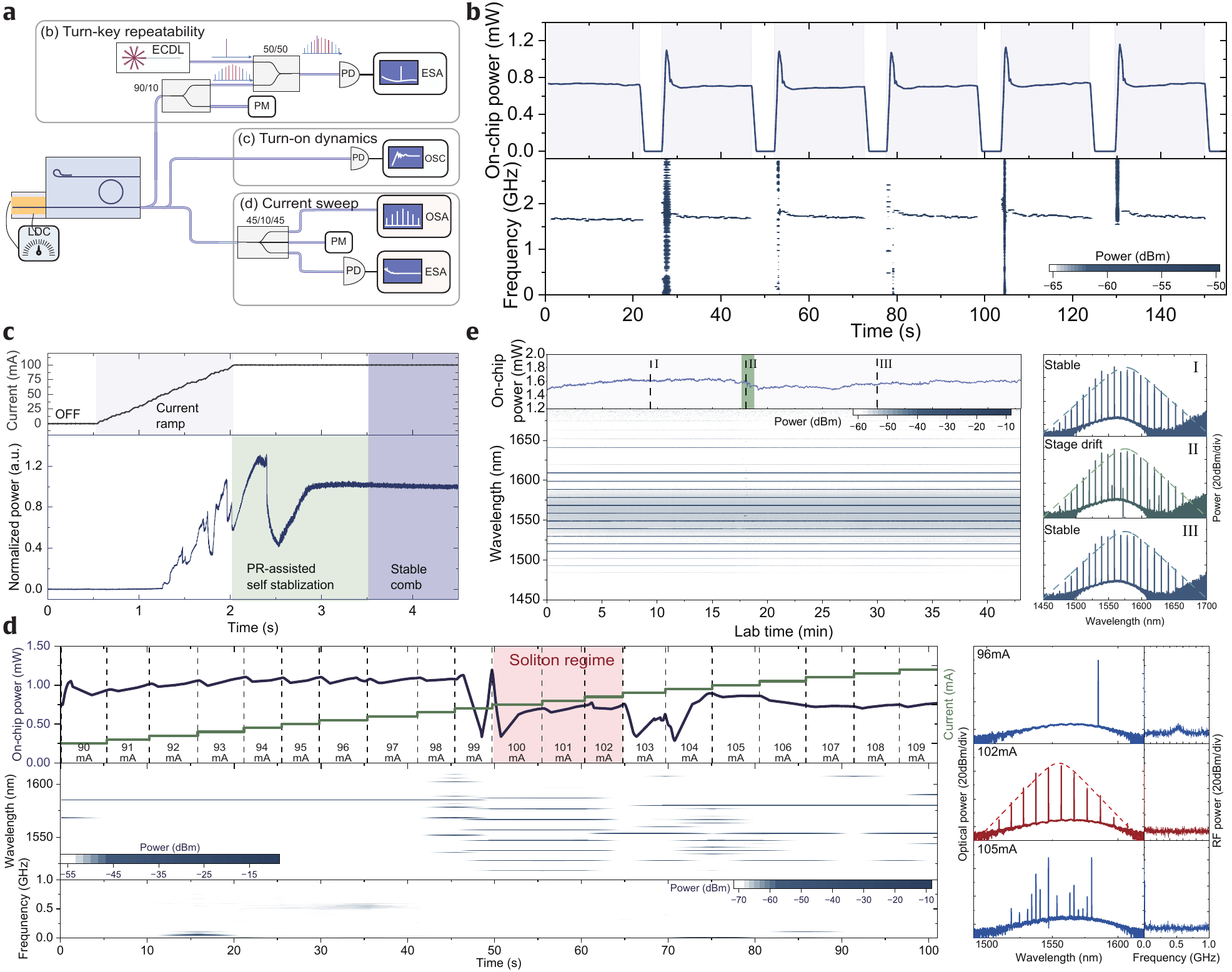}

\caption{\textbf{Dynamics and stability of turn-key operation.} 
\textbf{(a)}~Experimental setup to characterize the turn-key operation.
\textbf{(b)}~Time-dependent laser power (top) and beat-note spectrum (bottom) during five consecutive ON/OFF cycles (each with 20-s ON and 5-s OFF) of the RSOA current. Consistent recovery of both optical power and RF beatnote frequency confirms deterministic recurrence of the exact same soliton state. \textbf{(c)}~Startup dynamics of the laser power.
The dynamic process involves three phases: current ramp-up (grey), photorefraction-mediated transition (green), and steady-state soliton operation (purple). \textbf{(d)}~Laser power (top), optical (middle) and RF (bottom) spectra as the RSOA current is ramped from 90 to 109\,mA in 1-mA steps and $\sim$5-s duration per step. Representative spectra (right) show laser states before, during, and after the soliton regime. The lasing state becomes unstable when current $>$103\,mA, so the optical and RF spectra shown in this region only repsent a snapshot of certain lasing states. See Fig.\,S1 for detailed discussion and detailed spectra at each current step. \textbf{(e)}~Time-dependent laser power (top) and optical spectrum (bottom) over a long time duration, showing the long-term stability of the laser operation. Mechanical drift of the testing station perturbs the soliton operation after $\sim$18 minutes. A simple realignment of the mechanical stages readily restores the soliton state, indicating a robust mode-locking operation. The soliton state here exhibits a 3-dB optical bandwidth of 3.9\,THz (corresponding to a transform-limited pulse duration of $\sim$80\,fs) with a repetition rate of 1.2\,THz. It was obtained with optimized gain at a RSOA current of 160\,mA.}
\label{Fig3}
\end{figure*}

\section{Turn-key operation}
We investigated the turn-key capabilities of the soliton laser, examining both the reliability of soliton self-creation and the underlying temporal dynamics governing the initialization process. We performed multiple on‑off cycling while simultaneously monitoring both the comb power and the heterodyne beatnote with an external reference CW laser, as shown in Fig.\,\ref{Fig3}a. To ensure the beatnote accurately represented the comb state, we selected an off-center comb line (wavelength of 1550~nm) for the 1.2-THz soliton state at 100\,mA for heterodyning. As shown in Fig.\,\ref{Fig3}b, the beatnote consistently recurs at an identical frequency and the output optical power returns to the same level across multiple startup and shutdown cycles. This repeatable, ``set‑and‑forget" operation demonstrates the reliability of the soliton microcomb laser.

Notably, during each startup ramp, we observe a transient overshoot in comb power and a brief chaotic beatnote before the system settles into the steady soliton comb state. To probe these fast dynamics, we recorded the detailed output power waveform during a current ramp. During the current ramp from 0.5\,s to 2\,s, lasing commences at a RSOA current of $\sim50$\,mA, followed by several distinct power drop steps representing transitions between cavity lasing states (Fig.\,\ref{Fig3}c). After reaching the 100\,mA setpoint, the temporal trace exhibits larger, slower fluctuations before stabilizing at the characteristic power level of the soliton comb. The power variation displays a time constant of about tens to several hundred milliseconds, which is in agreement with our previous measurements of the PR effect on the TFLN platform\cite{jiang2017fasta}. This temporal behavior further confirms the pivotal role of photorefraction in facilitating reliable self‑starting operation.

To detail the lasing dynamics, we finely investigating the 1.2-THz soliton state at 100\,mA (Fig.\,\ref{Fig3}d). Below 99\,mA, the laser operates in a single-frequency continuous-wave (CW) state at $\sim$1585\,nm, with an on-chip output power of $\sim$1\,mW. This long-wavelength operation is likely favored because the microring's external coupling is stronger at longer wavelengths, resulting in lower intracavity loss at the red side of the C-band gain. At the 99\,mA threshold, the laser dynamics change abruptly. The output power briefly spikes, then drops sharply before settling at $\sim$0.75\,mW. Simultaneously, the optical spectrum transitions from a single CW line to a broadband, equally-spaced comb centered at $\sim$1560\,nm, which exhibits the ideal $sech^2$ envelope characteristic of a soliton. This transition is driven by the PR effect: as the RSOA current and intracavity power increase, the PR effect progressively blue-shifts the microresonator resonances. At 99\,mA, some FP cavity modes achieves the optimal red-detuning relative to microresonator resonances, enabling and initiating the sudden transition into the mode-locked soliton state. The observed power dynamics—a transient spike followed by a stable drop—are a classic signature of the system crossing the soliton existence threshold. This mode-locked soliton comb state remains stable as the current increases to $\sim$103\,mA. Beyond this point, the lasing becomes unstable. We attribute this to excessive detuning driven by the strengthened PR effect, pushing the laser away from the stable soliton regime. The detailed evolution of the optical and RF spectra during this transition is provided in the SI.

The soliton state remains very stable as soon as it is turned on. Fig.\,\ref{Fig3}e shows an example for a 1.2-THz soliton state, which exhibits a broad 3-dB bandwidth of 3.9\,THz (corresponding to a pulse width of $\sim$80\,fs). The state remains stable over the entire recording time period of about 46 minutes. As shown in Fig.\,\ref{Fig3}e-II, it is perturbed only by the mechanical drift of the testing station (as the RSOA gain chip and the TFLN PIC are placed on separate mechanical stages as shown in Fig.\,\ref{Fig1}d). The soliton state is fully recovered after the small mechanical drift is corrected manually (Fig.\,\ref{Fig3}e-III).  

\section{Ultralow comb linewidth}

\begin{figure}[htbp]
\centering
\includegraphics[width=0.5\textwidth]{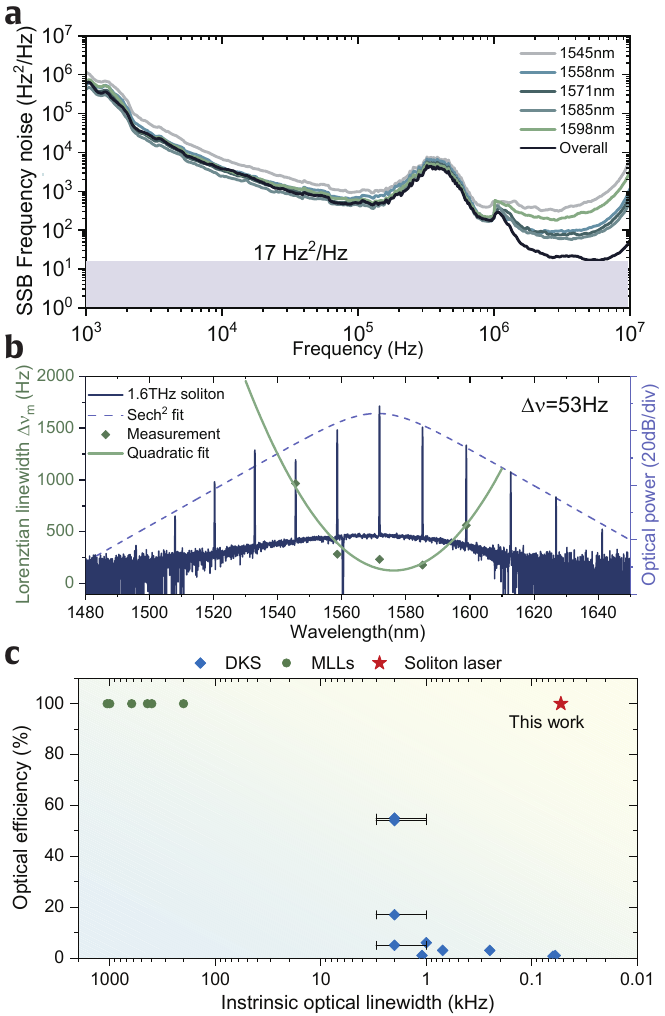}
\caption{\textbf{Optical linewidth of the soliton microcomb laser.} 
\textbf{(a)}~Frequency noise spectrum of individual comb lines and the overall comb for the 1.6-THz soliton state. The overall comb has frequency noise floor lower than individual comb lines likely because of detector noises that limit the characterizaton of the latter due to their lower powers. 
\textbf{(b)}~Spectral distribution of intrinsic optical linewidth for individual comb lines $\Delta \nu_m$, along with the soliton comb spectrum. 
\textbf{(c)}~Comparison of intrinsic optical linewidth and optical efficiency across different on-chip ultrafast comb sources. Complete performance metrics and definitions are provided in Table\,S2 of SI.}
\label{Fig4}
\end{figure}

Another significant challenge hindering the deployment of conventional on-chip MLLs is their limited coherence, a property that is critical for applications in communications, sensing, spectroscopy, and metrology. Most existing on-chip MLLs exhibit intrinsic optical linewidths in the sub-MHz range range, a limitation largely attributed to the high optical loss inherent to III-V materials\cite{davenport2018integratedb,wei2022advancesa}. To quantify the intrinsic linewidth of the soliton laser, we employed the correlated self‑heterodyne technique to measure the Lorentzian linewidth of individual comb teeth as well as the overall comb for the 1.6-THz soliton comb state at 120\,mA (Fig.\,S3). 

Fig.\,\ref{Fig4} shows the recorded frequency noise spectra and corresponding comb linewidths. 
As shown in Fig.\,\ref{Fig4}a, the overall soliton comb exhibits a white frequency noise floor of 17\,${\rm Hz^2/Hz}$, corresponding to an intrinsic Lorentzian linewidth as low as 53\,Hz. This value represents an improvement of at least three orders of magnitude over the best current chip-scale MLLs (Fig.\,\ref{Fig4}c) and is comparable to advanced on-chip free-running CW lasers\cite{margalit2021perspective}.
The superior linewidth performance of the soliton laser results directly from the nature of ultralow-loss TFLN external laser cavity structure that combines elegantly with an embedded high-Q microresonator for efficient mode locking. 

The intrinsic linewidth distribution across the spectrum exhibits a distinctive quadratic profile (Fig.\,\ref{Fig4}b). This parabolic dependence of individual comb-line linewidths further corroborates the mode-locked soliton nature, specifically the `elastic-tape' behavior characteristic of passively locked solitons\cite{takushima2004linewidth, kim2016ultralownoise}. Because comb modes are grown via FWM symmetrically around the center, consequently, comb lines further away from the center are more strongly affected by repetition rate jitter, leading to a squared dependence of linewidth on the mode number. 
Fig.\,\ref{Fig4}b shows that the spectral distribution of comb tooth linewidths displays an asymmetry about the central mode. The 1598\,nm tooth demonstrates a significantly narrower Lorentzian linewidth compared to the 1540\,nm mode, despite both being equidistant ($m=\pm2$) from the central mode. We attribute this asymmetry to stimulated Raman scattering (SRS) within the soliton, a phenomenon previously observed in DKS microcomb systems \cite{lei2022opticala}.

\section*{Discussion\label{sec:disc}}

To conclude, we have demonstrated the first on-chip femtosecond laser, enabled by photorefraction-assisted soliton mode-locking. This device achieves pulse durations shorter than 90\,fs with repetition rates up to 3\,THz, while simultaneously exhibiting an intrinsic linewidth of 53\,Hz. Our approach dramatically simplifies on-chip bright-soliton generation, offering stable turn-key operation, near-unity optical efficiency, superior long-term stability, and a remarkably low generation threshold (1\,V, 75\,mA) compatible with a standard alkaline battery. This soliton laser elegantly synthesizes the superior pulse duration and coherence of DKS microcombs with the operational simplicity and high efficiency of conventional on-chip MLLs (Fig.\,\ref{Fig4}c). The demonstrated soliton laser represents a fundamental advance in chip-scale frequency comb and pulse source technology, providing a compact and efficient solution that unlocks femtosecond pulse systems from controlled laboratory environments for diverse, real-world applications.

In the future, further integration with electrical tuning could possibly help fundamental mode-locking and high-speed soliton modulation\cite{he2023highspeedb}. By leveraging the rich optical properties of TFLN, particularly its quadratic nonlinearity and Pockels effect\cite{li2022integratedb,ling2024electricallyc}, this work opens a significant avenue toward the full on-chip integration of soliton comb lasers with multiple electro-optic and frequency conversion functionalities. This could accelerate various applications, such as self-referencing, frequency synthesis, high-speed photonic signal processing, and comb waveband conversion, all to be directly integrated on a single chip.

\section*{\label{sec:methods}Methods}

\subsection*{\label{sec:fab}Device Fabrication}
The devices were fabricated on a commercial z-cut lithium niobate on insulator (LNOI) wafer with a 600-nm-thick device layer. Waveguide patterns were defined using electron-beam lithography (JEOL) and subsequently transferred to the LNOI device layer via an argon ion milling process, etching to a depth of 420\,nm. Following the etch, the E-beam resist and any redeposited material were removed using a wet chemical cleaning process. To achieve coarse tuning of the FP laser cavity resonances, a set of devices with slightly varying round-trip length were fabricated on the same chip. After fabrication, the chip was diced and their facets were polished to minimize coupling losses to both the RSOA and the output lensed fiber. The chip-to-fiber coupling loss was measured to be $\sim$6\,dB per facet.

To favor unidirectional soliton excitation, the two coupling ports of the microring were designed with asymmetric coupling strengths. The port closer to the RSOA was designed to be over-coupled to ensure sufficient intracavity power, while the port on the Sagnac mirror side was designed to be critically coupled.

\subsection*{\label{sec:char}Laser Characterization}
For laser operation, the RSOA (Thorlabs) was mounted on a heatsink, and the TFLN chip was secured on a separate mechanical stage; neither component was actively temperature-controlled. The optical alignment between the RSOA, the TFLN chip, and a lensed fiber was first optimized to minimize coupling loss. The RSOA injection current was then swept to identify the operational range for soliton mode-locking. The laser output was collected from the chip facet using the lensed fiber.

Characterization of turn-key operation, lasing state evolution, and long-term stability shown in Fig.\,\ref{Fig3}b,c,e were performed by programmed coordinated tuning of RSOA injection current and simultaneously recording the optical spectrum, electrical spectrum, and laser power with an OSA (Yokogawa), ESA (Rhode Schwarz), and an optical power meter (Newport), respectively. The acquisition times for the ESA, OSA, and power meter were approximately 0.5\,s, 2\,s, and 0.25\,s, respectively. For the transient measurement shown in Fig.\,\ref{Fig3}c, the injection current was monitored by measuring the voltage drop across a low-resistance sensing resistor placed in series with the gain chip. This voltage signal was recorded simultaneously with the output of a fast photodetector detecting the soliton laser output.

Prior to pulse characterization with a frequency-resolved optical gating (FROG) instrument, the laser output was sent through a dispersion compensating fiber (DCF) and then amplified by a C-band EDFA (Amonics). The finite bandwidth and non-flat gain profile of the EDFA lead to non-uniform amplification of the comb lines, a known effect for THz-repetition-rate solitons that can distort the measured pulse shape. A detailed analysis is provided in the SI.

\subsection*{\label{sec:linewidth}Linewidth Measurement}
The fundamental linewidth was measured using a delayed self-heterodyne technique. The setup utilized a 17-m fiber delay line and an acousto-optic modulator (AOM) to generate a frequency-shifted beat note. The single-sided phase noise spectrum of the detected beat signal was measured with a phase noise analyzer (Keysight). The frequency noise spectral density was then calculated from the phase noise spectrum to determine the intrinsic linewidth. A detailed schematic of the setup is shown in SI, Fig.\,S3.

\section*{\label{sec:acknowledgement}Acknowledgements}
We thank Dr. Daniel Kane of MesaPhotonics for helpful discussions about the FROG traces. We also thank Dr. Wuxiucheng Wang and Yongchao Liu for their help with setup and current measurement. We thank Dr. Xiyuan Lu from NIST for the discussion about this work. This work is supported in part by the Defense Advanced Research Projects Agency (DARPA) NaPSAC program (Agreement No.~N660012424007) and the National Science Foundation (NSF) (OMA-2138174, ECCS-2231036, OSI-2329017). This work was performed in part at the Cornell NanoScale Facility, a member of the National Nanotechnology Coordinated Infrastructure (NNCI), which is supported by the National Science Foundation (Grant NNCI-2025233).
\section*{\label{sec:author clare}Author contributions}
Q.H., J.L., and Z.G. designed the devices. Q.H. fabricated the devices. Q.H. performed all device characterization. R.L., S.X., and J.S. assisted in the experiments. Y.H. provided guidance on the DKS experiment. Q.H. and Q.L. wrote the manuscript with contributions from all authors. Q.L. supervised the project. Q.L. conceived the concept.

\bibliography{Ref.bib}
\bibliographystyle{naturemag}

\end{document}